# TDMA Achieves the Optimal Diversity Gain in Relay-Assisted Cellular Networks

Suzhi Bi, Ying Jun (Angela) Zhang, *Senior Member, IEEE*

Department of Information Engineering, The Chinese University of Hong Kong,

Shatin, New Territories, Hong Kong.

Email: {bsz009, yjzhang}@ie.cuhk.edu.hk

## Abstract

In multi-access wireless networks, transmission scheduling is a key component that determines the efficiency and fairness of wireless spectrum allocation. At one extreme, greedy opportunistic scheduling that allocates airtime to the user with the largest instantaneous channel gain achieves the optimal spectrum efficiency and transmission reliability but the poorest user-level fairness. At the other extreme, fixed TDMA scheduling achieves the fairest airtime allocation but the lowest spectrum efficiency and transmission reliability. To balance the two competing objectives, extensive research efforts have been spent on designing opportunistic scheduling schemes that reach certain tradeoff points between the two extremes. In this paper and in contrast to the conventional wisdom, we find that in relay-assisted cellular networks, fixed TDMA achieves the same optimal diversity gain as greedy opportunistic scheduling. In addition, by incorporating very limited opportunism, a simple relaxed-TDMA scheme asymptotically achieves the same optimal system reliability in terms of outage probability as greedy opportunistic scheduling. This reveals a surprising fact: transmission reliability and user fairness are no longer contradicting each other in relay-assisted systems. They can be both achieved by the simple TDMA schemes. For practical implementations, we further propose a fully distributed algorithm to implement the relaxed-TDMA scheme. Our results here may find applications in the design of next-generation wireless communication systems with relay architectures such as LTE-advanced and WiMAX.

This work was supported in part by the Competitive Earmarked Research Grant (Project Number 419509) established under the University Grant Committee of Hong Kong.

# I. INTRODUCTION

*A. Motivations and Summary of Contributions*

Relay-assisted transmission techniques are known to be effective in combating path loss and enhancing link quality in wireless communications systems [1,2,3]. Such techniques are already adopted in the 4G wireless communications standards, such as LTE-Advanced and IEEE WiMAX [4,5]. In these systems, fixed relays are deployed as intermediate nodes to forward data between mobile users and base stations (BS), thus extending the service coverage of a cell and enhancing the overall throughput performance of the system.

In multi-access wireless systems, transmission scheduling is a key component that determines the efficiency and fairness of spectrum resource allocation. In particular, opportunistic scheduling that takes advantage of independent time-varying channels across different mobile users effectively exploits multiuser diversity through scheduling the transmission of users according to their instantaneous channel conditions. Depending on their greediness, different opportunistic scheduling schemes achieve different tradeoffs between user-level fairness and high system performance in terms of throughput or transmission reliability. At one extreme, greedy opportunistic scheduling, which selects the user with the largest instantaneous channel capacity, achieves the highest spectrum efficiency and transmission reliability but the worst fairness among users. On the other extreme, being oblivious to channel states, fixed TDMA achieves fairest airtime allocation but lowest transmission reliability. To balance the two competing objectives, different opportunistic scheduling policies have been designed to reach certain tradeoff between the two extremes [7,8,9,10,11]. One such example is proportional fair scheduling, which schedules transmissions according to the users' "relative" channel strengths [8]. Less greedier than greedy opportunistic scheduling, proportional fair scheduling achieves equal airtime allocation among users in the long run. Another interesting work uses an $\alpha$-Rule scheduling to achieve a flexible tradeoff between spectrum efficiency and user fairness by tuning the variable $\alpha$ in its scheduling policy [9].

In this paper and in contrast to conventional belief, we find that the transmission reliability and user fairness are no longer contradicting each other in relay-assisted cellular systems. With optimal relay selection, fixed TDMA achieves the same optimal diversity gain as greedy opportunistic scheduling. In addition, by incorporating very limited opportunism, a simple relaxed-TDMA scheme asymptotically achieves the same optimal outage probability as greedy opportunistic scheduling. In other words, we can fully enjoy the multiuser diversity gain achievable by greedy opportunistic scheduling without suffering its disadvantages such as poor user fairness and high implementation cost. Our contributions are detailed below.

- We derive the optimal outage probability in relay-assisted cellular networks. In particular, we show that the optimal outage probability is achieved by greedy opportunistic scheduling, which fully exploits multiuser diversity. By letting the number of users go to infinity, a lower bound on the outage probability is obtained. Interestingly, this lower bound is independent of the user-side parameters including the transmission powers and channel conditions of users.

- We find that fixed TDMA scheduling achieves the same diversity gain as greedy opportunistic scheduling in relay-assisted networks. In addition, we show that the diversity order is solely determined by the number of relays. This implies that the greediness of the scheduling policy is irrelevant when it comes to diversity order in relay-assisted cellular network. This is quite different from most wireless systems where the diversity order is closely related to the scheduling policy.

- A power gap is observed between fixed TDMA and greedy opportunistic scheduling. We quantify this power gap in high SNR region and find that the gap depends on the power allocation ratio between mobile users and relay set. Interestingly, we show that this power gap is closed when minimum opportunism is introduced to TDMA. In particular, a relaxed-TDMA scheme asymptotically achieves the optimal outage probability that is otherwise achievable by greedy opportunistic scheduling. This reveals an encouraging fact: through the use of TDMA, optimal outage probability can be achieved in relay-assisted networks without compromising the user-level fairness.

- We propose a fully distributed algorithm to implement the relaxed-TDMA scheme, where the scheduling decision is made in a distributed manner at the relays based on their own local channel conditions. We show that, the proposed distributed algorithm achieves the optimal outage probability while generating very little signaling overhead.

The rest of this paper is organized as follows. We describe the system model and the optimal relay selection method in Section II. In Section III, we derive the explicit expression of optimal outage probability in relay-assisted networks. It is proved in Section IV that fixed TDMA yields the same diversity gain as greedy opportunistic scheduling. A relaxed-TDMA scheme is introduced in Section V, where we show that it achieves the optimal outage probability and high user-level fairness at the same time. Simulations results are given in Section VI. Finally, the paper is concluded in Section VII.

*B. Related Works*

Like in traditional cellular networks, channel-aware opportunistic user scheduling is of great interests in relay-assisted cellular networks. A common objective shared by most recent work is to maximize the total throughput [12,13,14,15]. This, however, leads to poor user-level fairness, for some users may experience excessively long access delay, if they are stuck in deep fading channels. With this in mind, some work aims to strike a balance between spectrum efficiency and fairness by applying scheduling schemes that are less aggressive [3,16,17]. For example, [3] incorporated the queue-size information into its scheduling protocol design, where the downlink throughput is maximized under the constraint that queue-length at all nodes are finite. [17] extended proportional fair scheduling to relay-assisted systems to achieve long-term user-level fairness at the cost of throughput reduction. In contrast to conventional wisdom, our work shows that with optimal relay selection, a simple relaxed-TDMA scheduling scheme obtains the optimal outage probability and excellent fairness among users at the same time. This indicates that the conventional tradeoff between transmission reliability and user fairness is not necessary in relay-assisted cellular networks.

Opportunistic scheduling in relay-assisted networks normally requires strong centralized control at the BS to coordinate the transmissions of both mobile users and relays. In [3,12,13], the channel between every two nodes is estimated and fed back to the BS for centralized processing. The cost of either time or bandwidth on transmitting large amount of pilot signals inevitably decreases the overall spectrum efficiency. On the other hand, [18,19] proposed distributed scheduling schemes that allow the relays to participate in scheduling decisions based on the limited local channel state information (CSI) at the relays. Compared with centralized schemes, distributed implementations effectively reduce signaling overhead and processing delay. Nevertheless, the proposed distributed schemes decouple user scheduling and relay selection process, thus are suboptimal compared to centralized scheduling schemes. By contrast, our distributed scheduling protocol achieves optimal outage probability through jointly scheduling user and selecting the relay.

When a user is scheduled, a proper set of relays needs to be assigned to assist its transmission. The optimal strategy that maximizes the received SNR at the BS is for all available relays to form a virtual antenna array and jointly transmit the source information using beamforming techniques [1]. However, beamforming is costly to implement in distributed relay networks, since it requires the knowledge of global CSI and strong centralized coordination. Similarly, a distributed space-time code (DSTC) scheme that makes use of all available relays is proposed by Laneman [20]. Although DSTC achieves full diversity gain, practical distributed space-time code design is very difficult, since the set of available relays is time varying due to channel fading. Besides, its implementation requires strict symbol level synchronization, which is also considered difficult in distributed networks. Alternatively, single-relay selection, which employs only one "best" relay to assist transmission, greatly reduces system implementation complexity and saves significant signaling overhead. Besides, recent studies also show that single-relay selection schemes achieve comparable system reliability as multi-relay transmission schemes. For example, [22,23] showed that single-relay selection schemes yield near optimal outage performance. [21] proved that single-relay selection achieves lowest outage probability under aggregate relay power constraint. [24] further showed that single-relay selection outperforms DSTC scheme when the number of relays is greater than three. In this paper, we will also demonstrate the optimality of the single-relay selection method.

Before leaving the session, we would like to emphasize that relay selection is viewed as part of the operations at the network infrastructure side. For each scheduled user, there exists a mechanism that assigns a proper set of relays to assist the transmission of the user. In this paper, our focus is on the scheduling of the users, given that an optimal relay selection mechanism is used.

## II. SYSTEM MODEL AND OPTIMAL RELAY SELECTION METHOD

### A. System Model

We consider the uplink of a cellular network with a base station (BS), $N$ stationary relays and $M$ mobile users communicating to the BS through the relays. The direct user-to-BS links are assumed to be non-existent, so that all user-to-BS communicationstake place in a two-hop manner through the relays. Each relay works in a half-duplex mode using decode and forward (DaF) scheme. Suppose that all users and relays transmit with a fixed data rate. The received message can be correctly decoded only when the received signal-to-noise ratio (SNR) exceeds a prescribed threshold $\tau$.

Suppose that channel fading is independent across different links. Moreover, channel fading is assumed to remain unchanged during each two-hop transmission period. In the first hop, a mobile user, say user $u$, is scheduled to transmit its signal $x_u$ with power $P_{user}$, where $E[|x_u|]^2 = 1$. Then, the received signal at the $r^{th}$ relay is

$$y_r = \sqrt{P_{user}} \alpha_{u,r} x_u + n_r, \qquad r = 1,2,\ldots N. \tag{1}$$

Here, $\alpha_{u,r} \sim \mathcal{CN}(0, \Omega_{u,r})$ is the instantaneous channel fading coefficient between the $u^{th}$ user and the $r^{th}$ relay. Denoted by $\gamma_{u,r} \triangleq |\alpha_{u,r}|^2$, the instantaneous channel gain follows an exponential distribution with mean $E[\gamma_{u,r}] = \Omega_{u,r}$. The noise $n_r$ is assumed to be i.i.d and $n_r \sim \mathcal{CN}(0, N_0)$. Then the received SNR at the $r^{th}$ relay is $P_{user}\gamma_{u,r}/N_0$. Let $\mathcal{D}_u$ be the set of relays that successfully decode the $u^{th}$ mobile user's message. That is, relay $r \in \mathcal{D}_u$ if $P_{user}\gamma_{u,r}/N_0 \geq \tau$. The relays in $\mathcal{D}_u$ are referred to as decoding relays.

In the second hop, all or a subset of decoding relays forward $x_u$ to the BS using different orthogonal channels, e.g. in separate frequency channels, to avoid mutual interference. At first glance, this orthogonal transmission scheme requires excessive bandwidth to accommodate numerous relays.

Fortunately, as we will show in subsection II-B, it is optimal to allow only one decoding relay to transmit in the second hop. Some careful readers may suggest using multiple relays to form an optimal distributed antenna array that maximizes the received SNR at the BS. It is, however, costly to implement this method. This is mainly because to form the transmission beamforming vector, each relay will have to know the indices of all the decoding relays as well as the channel conditions from these relays to the BS[1]. In contrast, as we will show in a later section, our proposed method can be implemented in a fully distributed manner where each relay only needs to know the channel fading of its own link.

Suppose that the $r^{th}$ relay transmits with power $P_r$, and $\sum_{r \in \mathcal{D}_u} P_r = P_{relay}$, where $P_{relay}$ is the total transmission power allocated to the relay period. Then, the received signal at the BS is

$$z_r = \sqrt{P_r} \alpha_{r,B} x_u + n_{r,B}, \qquad r \in \mathcal{D}_u. \tag{2}$$

To maximize the output SNR, the BS combines the received signal using maximum ratio combination as follows,

$$\begin{aligned} z_B &= \sum_{r \in \mathcal{D}_u} \sqrt{P_r}\, \alpha_{r,B}^{*} z_r \\ &= \left( \sum_{r \in \mathcal{D}_u} P_r\, \gamma_{r,B} \right) x_u + \sum_{r \in \mathcal{D}_u} \sqrt{P_r}\, \alpha_{r,B}^{*} n_{r,B}, \end{aligned} \tag{3}$$

where $\gamma_{r,B} = |\alpha_{r,B}|^2$ are the instantaneous channel gains of the relay-BS (R-B) channels. The corresponding maximum received SNR is calculated as

$$SNR_B = \frac{\left(\sum_{r \in \mathcal{D}_u} P_r\, \gamma_{r,B}\right)^2 E[x_u^2]}{N_0 \sum_{r \in \mathcal{D}_u} P_r\, \gamma_{r,B}} = \frac{\sum_{r \in \mathcal{D}_u} P_r\, \gamma_{r,B}}{N_0}. \tag{4}$$

An outage event occurs when $SNR_B < \tau$. The probability of such an event, referred to as outage probability, is a key metric to measure the transmission reliability of the system.

## B. Relay Selection method

The following proposition shows that selecting a single relay obtains the highest $SNR_B$ and hence the lowest outage probability.

**Proposition 1**: Selecting the relay $r^* \in \mathcal{D}_u$ with the largest R-B channel gain yields the lowest outage probability.

*Proof:* We can infer from (4) that

---

[1] The optimal beamforming is the maximum ratio transmission (MRT) beamforming, where the relay $r \in \mathcal{D}_u$ transmits $\frac{\sqrt{P_{relay}} \alpha_{r,B}^{*}}{\sqrt{\sum_{m \in \mathcal{D}_u} |pha_{m,B}|^2}} x_u$ to the BS. The calculation of the beamforming vector clearly requires the knowledge of both indices and channel state information of all decoding relays.

$$SNR_B = \sum_{r \in \mathcal{D}_u} \gamma_{r,B} \frac{P_r}{N_0} \leq \sum_{r \in \mathcal{D}_u} \left(\max_{r \in \mathcal{D}_u} \gamma_{r,B}\right) \frac{P_r}{N_0} = \left(\max_{r \in \mathcal{D}_u} \gamma_{r,B}\right) \frac{P_{relay}}{N_0}. \quad (5)$$

This shows that the highest SNR, and hence the lowest outage probability, is obtained by allocating all transmission power to the relay with the highest instantaneous R-B channel gain. That is, it is optimal to let only one "best" relay to relay the message, where the "best" relay is chosen as

$$r_u^* = \max_{r \in \mathcal{D}_u} \gamma_{r,B}. \quad (6)$$

∎

Based on the above "single-relay selection" argument, we propose in Corollary 1 the optimal relay selection scheme for relay-assisted cellular networks. This scheme is convenient for both analysis and distributed implementation. Note that the the scheme in Corollary 1 is optimal in the sense that it yields the same optimal outage probability as the one in Proposition 1. Nonetheless, it is possible that the two schemes end up selecting different but equally optimal relays.

**Corollary 1**: For the $u^{th}$ mobile user, allocating full transmission power to the "best" relay $r_u^*$, given by

$$r_u^* = \arg \max_{r=1,2,..,N} \min \{P_{user}\gamma_{u,r}, P_{relay}\gamma_{r,B}\}, \quad (7)$$

yields lowest outage probability.

**Remark 1**: (7) is indeed selecting the relay that provides the best end-to-end channel.

*Proof of Corollary 1:* Let us refer to the optimal relay selection method in (5) as Method $\theta$, and in (7) as Method $\pi$. Define a set $\mathcal{S}$ where each entry $\mathbf{x}_u \in \mathcal{S}$ is a $2N \times 1$ channel statevector for the $u^{th}$ user, i.e. $\mathbf{x}_u = [\gamma_{u,r}, \gamma_{r,B}]'$ and $r = 1,2,..,N$. We say $\mathbf{x}_u \in \mathcal{S}_\theta^{out}$, if selecting a relay using Method $\theta$ under channel state $\mathbf{x}_u$ yields an outage. The complement of $\mathcal{S}_\theta^{out}$, denoted as $\overline{\mathcal{S}_\theta^{out}}$ contains the channel states that result in a successful transmission. Similarly, we define $\mathcal{S}_\pi^{out}$ and $\overline{\mathcal{S}_\pi^{out}}$ for Method $\pi$. According to (6),we note that an outage occurs in Method $\pi$ when

$$\max_{r=1,2,..,N} \min \{P_{user}\gamma_{u,r}, P_{relay}\gamma_{r,B}\} < \tau N_0. \quad (8)$$

To prove Corollary 1, all we need to show is that $\mathcal{S}_\theta^{out} = \mathcal{S}_\pi^{out}$.

Suppose that a channel state vector $\mathbf{x}_u^0 = \{\gamma_{u,r}^0, \gamma_{r,B}^0\} \in \mathcal{S}_\theta^{out}$. In this case, for those relays $r \in \mathcal{D}_u$, we have $P_{relay}\gamma_{r,B}^0 < \tau N_0$. And for the relays $r$ that are not in $\mathcal{D}_u$, $P_{user}\gamma_{u,r}^0 < \tau N_0$ holds by definition. Therefore, the following inequality holds for all relays

$$\min \{P_{user}\gamma_{u,r}^0, P_{relay}\gamma_{r,B}^0\} < \tau N_0, r = 1,2,..,N. \quad (9)$$

This is equivalent to

$$\max_{r=1,2,..,N} \min \{P_{user}\gamma_{u,r}^0, P_{relay}\gamma_{r,B}^0\} < \tau N_0. \quad (10)$$

Note that when the above inequality holds, an outage also occurs with Method $\pi$, i.e. $\boldsymbol{x}^0 \in \mathcal{S}_\pi^{out}$. Therefore, we have $\mathcal{S}_\theta^{out} \subseteq \mathcal{S}_\pi^{out}$.

Next, we show that $\mathcal{S}_\pi^{out} \subseteq \mathcal{S}_\theta^{out}$. Let us consider a channel state vector $\boldsymbol{x}_u^1 = \{\gamma_{u,r}^1, \gamma_{r,B}^1\} \in \overline{\mathcal{S}_\theta^{out}}$. Then there must be at least one relay $r'$ that satisfies both $P_{user}\gamma_{u,r'}^1 \geq \tau N_0$ and $P_{relay}\gamma_{r',B}^1 \geq \tau N_0$, or

$$\min \{P_{user}\gamma_{u,r'}^1, P_{relay}\gamma_{r',B}^1\} \geq \tau N_0. \quad (11)$$

That is to say

$$\max_{r=1,2,..,N} \min \{P_{user}\gamma_{u,r}^1, P_{relay}\gamma_{r,B}^1\} \geq \tau N_0, \quad (12)$$

which means $\boldsymbol{x}^1 \in \overline{\mathcal{S}_\pi^{out}}$ too. Therefore we have $\overline{\mathcal{S}_\theta^{out}} \subseteq \overline{\mathcal{S}_\pi^{out}}$, or equivalently $\mathcal{S}_\pi^{out} \subseteq \mathcal{S}_\theta^{out}$.

Now that $\mathcal{S}_\theta^{out} \subseteq \mathcal{S}_\pi^{out}$ and $\mathcal{S}_\pi^{out} \subseteq \mathcal{S}_\theta^{out}$ hold simultaneously, it is sufficient to claim that $\mathcal{S}_\theta^{out} = \mathcal{S}_\pi^{out}$. That is, both methods yield the same outage probability. Since Method $\theta$ yields the lowest outage probability according to Proposition 1, selecting the relay according to (7) is also outage optimal. ∎

So far, we have described the way to optimally select a single relay that minimizes the outage probability for any mobile user that is being scheduled to transmit. A distributed algorithm that implements this optimal relay selection scheme will be introduced in the Appendix. In the remainder of this paper, we will focus on transmission scheduling at the user side, which is the main concern of this paper.

## III. OPTIMAL OUTAGE PROBABILITY OF RELAY-ASSISTED NETWORKS

### A. Optimality of Greedy Opportunistic Scheduling

The key idea of greedy opportunistic scheduling is to allocate airtime to the mobile user with the largest instantaneous channel gain. With the optimal relay selection method, greedy opportunistic scheduling in a relay assisted network selects the "best" user $u^*$ according to

$$u^* = \arg \max_{u=1,2,..,M} \min \{P_{user}\gamma_{u,r_u^*}, P_{relay}\gamma_{r_u^*,B}\}, \quad (13)$$

where the optimal relay $r_u^*$ for the $u^{th}$ mobile user is given in (7).

The following theorem proves that greedy opportunistic scheduling yields the lowest outage probability, which is intuitive.

**Theorem 1**: In multi-access DaF relay-assisted networks with an aggregate relay power constraint,

greedy user scheduling in (13) is outage optimal.

*Proof:* Denote by $(u^*, r_{u^*}^*)$ the user-relay pair selected according to (13). Consider another user-relay pair $(u,r) \neq (u^*, r_{u^*}^*)$, where the inequality means the two equalities $u = u^*$ and $r = r_{u^*}^*$ do not hold simultaneously. As proved in Corollary 1, selecting $r_u^*$ according to (7) yields optimal outage probability for the particular $u$ under aggregate relay power constraint. That is

$$P_{out}[(u, r_u^*)] \leq P_{out}[(u,r)], \forall r. \tag{14}$$

Meanwhile, the selection in (1) guarantees that

$$\min\left\{P_{user}\gamma_{u^*,r_{u^*}^*}, P_{relay}\gamma_{r_{u^*}^*,B}\right\} \geq \min\left\{P_{user}\gamma_{u,r_u^*}, P_{relay}\gamma_{r_u^*,B}\right\}, \forall u. \tag{15}$$

Hence, $(u^*, r_{u^*}^*)$ yields lower outage probability than $(u, r_u^*)$. As a result, we have

$$P_{out}[(u^*, r_{u^*}^*)] \leq P_{out}[(u, r_u^*)] \leq P_{out}[(u,r)], \forall u, r. \tag{16}$$

This completes the proof. ∎

### 3.2. Outage Probability Analysis

In this subsection, we compute the optimal outage probability achieved by greedy opportunistic scheduling. By definition, we have

$$P_{out} = Pr\left\{\max_{u=1,2,\ldots,M} \max_{r=1,2,\ldots,N} \min\{P_{user}\gamma_{u,r}, P_{relay}\gamma_{r,B}\} < \tau N_0\right\}. \tag{17}$$

To simplify the notations, we define a decision parameter $W$ as follows

$$\begin{aligned} W &\triangleq \max_{u=1,2,\ldots,M} \max_{r=1,2,\ldots,N} \min\{P_{user}\gamma_{u,r}, P_{relay}\gamma_{r,B}\} \\ &= \max_{r=1,2,\ldots,N} \max_{u=1,2,\ldots,M} \min\{P_{user}\gamma_{u,r}, P_{relay}\gamma_{r,B}\} \\ &= \max_{r=1,2,\ldots,N} \min\left\{\max_{u=1,2,\ldots,M} P_{user}\gamma_{u,r}, P_{relay}\gamma_{r,B}\right\}. \end{aligned} \tag{18}$$

Then, $P_{out}$ can be written as

$$P_{out} = Pr[W < \tau N_0]. \tag{19}$$

Let $X_r \triangleq \max_{u=1,2,\ldots,M}(P_{user}\gamma_{u,r})$. The probability distribution of $X_r$ is

$$\begin{aligned} Pr[X_r < x] &= Pr\left[\max_{u=1,2,\ldots,M} \gamma_{u,r} < \frac{x}{P_{user}}\right] \\ &= \prod_{u=1}^{M} Pr\left[\gamma_{u,r} < \frac{x}{P_{user}}\right] \\ &= \prod_{u=1}^{M}\left(1 - \exp\left\{-\frac{x}{P_{user}\Omega_{u,r}}\right\}\right). \end{aligned} \tag{20}$$

Likewise, let $Y_r \triangleq \min\{X_r, P_{relay}\gamma_{r,B}\}$. Since $\gamma_{r,B}$ is independent of $X_r$ (user-to-relay (U-R) and R-B channels are independent), the probability distribution of $Y_r$ can be calculated as

$$\begin{aligned}
Pr[Y_r < x] &= 1 - Pr[Y_r \geq x] \\
&= 1 - Pr[X_r \geq x] \cdot Pr\left[\gamma_{r,B} \geq \frac{x}{P_{relay}}\right] \\
&= 1 - \left[1 - \prod_{u=1}^{M}\left(1 - \exp\left\{-\frac{x}{P_{user}\Omega_{u,r}}\right\}\right)\right] \cdot \exp\left\{-\frac{x}{P_{relay}\Omega_{r,B}}\right\}.
\end{aligned} \quad (21)$$

Since $Y_r$'s are independent for different $r$, we can obtain the system outage probability

$$\begin{aligned}
P_{out} &= Pr[W < \tau N_0] \\
&= Pr\left[\max_{r=1,..,N} Y_r < \tau N_0\right] \\
&= \prod_{r=1}^{N} Pr[Y_r < \tau N_0] \\
&= \prod_{r=1}^{N}\left\{1 - \left[1 - \prod_{u=1}^{M}\left(1 - \exp\left\{-\frac{\tau N_0}{P_{user}\Omega_{u,r}}\right\}\right)\right] \cdot \exp\left\{-\frac{\tau N_0}{P_{relay}\Omega_{r,B}}\right\}\right\}.
\end{aligned}$$
(22)

Define SNR by $\eta = \frac{P_0}{N_0}$, where $P_0 = P_{user} + P_{relay}$ is the total power needed to transmit one symbol. Let $P_{user} = \alpha P_0$ and $P_{relay} = (1 - \alpha)P_0$, where $\alpha \in (0,1)$. From (22), we can express the outage probability as a function of $M$, $N$ and SNR

$$P_{out}(M, N, \eta) = \prod_{r=1}^{N}\left\{1 - \left[1 - \prod_{u=1}^{M}\left(1 - \exp\left\{-\frac{\tau}{\alpha\eta\Omega_{u,r}}\right\}\right)\right] \cdot \exp\left\{-\frac{\tau}{(1-\alpha)\eta\Omega_{r,B}}\right\}\right\}.$$
(23)

It can be seen from (23) that the outage probability decreases as we increase either $M$ or $N$. In the extreme case when the number of users becomes very large, we can obtain the lower bound on the outage probability as follows

$$\begin{aligned}
\underline{P_{out}}(N, \eta) = P_{out}(\infty, N, \eta) &= \prod_{r=1}^{N}\left\{1 - \exp\left\{-\frac{\tau}{(1-\alpha)\eta\Omega_{r,B}}\right\}\right\} \\
&= \prod_{r=1}^{N}\left\{1 - \exp\left\{-\frac{\tau N_0}{P_{relay}\Omega_{r,B}}\right\}\right\}.
\end{aligned}$$
(24)

Note that the lower bound only depends on the parameters related to the relays, including the relay transmission power $P_{relay}$ and R-B mean channel gains $\Omega_{r,B}$. It is, however, not dependent on the user-side parameters such as $P_{user}$ and $\Omega_{u,r}$. Due to the optimality of greedy opportunistic scheduling, $\underline{P_{out}}$ is also the lowest outage probability in a DaF relay-assisted network.

*C. Diversity Order Analysis*

It is not surprising that $P_{out}$ decreases as we increase either $M$ or $N$ in (23), as more users or relays yield a higher order of diversity. However, in what follows, we show that diversity order only depends on the relay number $N$. In other words, applying greedy opportunistic scheduling among a large number of users, i.e. having large $M$, is immaterial in improving the outage performance at high SNR region.

Define diversity order as the negative slope of the outage probability as a function of SNR in a log-log plot, i.e.,

$$d = - \lim_{\eta \to \infty} \frac{\log P_{out}(M, N, \eta)}{\log(\eta)}. \tag{25}$$

Noting that $\exp\{-x\} \approx 1 - x$, as $x \to 0$ and $x > 0$, we can approximate the right hand side (RHS) of (23) in high SNR region by

$$P_{out}(M, N, \eta) \approx \prod_{r=1}^{N} \left\{ \frac{\tau}{(1-\alpha)\Omega_{r,B}} \cdot \frac{1}{\eta} + \prod_{u=1}^{M} \frac{\tau}{\alpha\Omega_{u,r}} \cdot \left(\frac{1}{\eta}\right)^M - \prod_{u=1}^{M} \frac{\tau}{\alpha\Omega_{u,r}} \cdot \frac{\tau}{(1-\alpha)\Omega_{r,B}} \cdot \left(\frac{1}{\eta}\right)^{M+1} \right\}. \tag{26}$$

When $\eta \to \infty$, (26) can be simplified as

$$\lim_{\eta \to \infty} P_{out}(M, N, \eta) = \begin{cases} \prod_{r=1}^{N} \left( \frac{\tau}{(1-\alpha)\Omega_{r,B}} + \frac{\tau}{\alpha\Omega_{1,j}} \right) \left(\frac{1}{\eta}\right)^N, & M = 1 \\ \prod_{r=1}^{N} \frac{\tau}{(1-\alpha)\Omega_{r,B}} \left(\frac{1}{\eta}\right)^N, & M > 1. \end{cases} \tag{27}$$

(27) shows that the diversity order $d$ is

$$d = - \lim_{\eta \to \infty} \frac{\log P_{out}(M, N, \eta)}{\log(\eta)} = N, \tag{28}$$

which implies that diversity order of outage probability here is dominated by the number of relays, regardless of the number of users involved in the greedy opportunistic scheduling. This is in contrast to opportunistic scheduling in conventional wireless systems where a larger number of users yields a higher diversity order. The result here provides a guideline in the system design: to decrease the outage probability, it is much more effective to increase the number of relays than to do large scale greedy opportunistic scheduling.

Note that the variable $M$ in the above equations is not necessarily the number of users in the cell. It is indeed the number of users among which opportunistic scheduling is applied, if opportunistic scheduling is performed in a smaller scale. For example, if the users are partitioned into groups and opportunistic scheduling is applied within each group, then this $M$ should be replaced by the group size.

## IV. FIXED TDMA ACHIEVES FULL DIVERSITY GAIN

It is commonly believed that to achieve the best system reliability, greedy opportunistic scheduling should be applied to select the "best" user to transmit every time. Indeed, this is consistent with our analysis above, which shows that optimal outage probability is obtained by greedy opportunistic scheduling. However, this optimal transmission reliability comes at the cost of the poor fairness among users. The greedy opportunistic scheduling are strongly biased to schedule the user with the best average U-R channel conditions. On the other hand, fixed TDMA achieves the fairest airtime allocation among users. However, it fails to exploit multiuser diversity and is commonly believed to achieve the lowest diversity order. In this section and in contrast to the common belief, we show that fixed TDMA achieves the same full diversity order as greedy opportunistic scheduling in relay-assisted networks, as long as the relay is selected properly.

### A. Diversity order of fixed TDMA

The outage probability of fixed TDMA can be calculated by letting $M = 1$ in (23). The $u^{th}$ user, for example, transmits with outage probability equal to

$$\prod_{r=1}^{N}\left\{1 - \exp\left\{-\frac{\tau}{\eta}\left(\frac{1}{\alpha\Omega_{u,r}} + \frac{1}{(1-\alpha)\Omega_{r,B}}\right)\right\}\right\}. \tag{29}$$

The outage probability of the system is an average of that of the $M$ individual users. Thus,

$$P_{out}^{TDMA}(M, N, \eta) = \frac{1}{M}\sum_{u=1}^{M}\left(\prod_{r=1}^{N}\left\{1 - exp\left\{-\frac{\tau}{\eta}\left(\frac{1}{\alpha\Omega_{u,r}} + \frac{1}{(1-\alpha)\Omega_{r,B}}\right)\right\}\right\}\right). \tag{30}$$

In the high SNR region, we can approximate $P_{out}^{TDMA}$ as

$$\begin{aligned}P_{out}^{TDMA}(M, N, \eta) &\approx \frac{1}{M}\sum_{u=1}^{M}\left(\prod_{r=1}^{N}\left\{\frac{\tau}{\eta}\left(\frac{1}{\alpha\Omega_{u,r}} + \frac{1}{(1-\alpha)\Omega_{r,B}}\right)\right\}\right) \\ &= \left\{\frac{1}{M}\sum_{u=1}^{M}\left(\prod_{r=1}^{N}\left\{\frac{\tau}{\alpha\Omega_{u,r}} + \frac{\tau}{(1-\alpha)\Omega_{r,B}}\right\}\right)\right\}\left(\frac{1}{\eta}\right)^{N},\end{aligned} \tag{31}$$

which leads to a diversity order of

$$d = -\lim_{\eta\to\infty}\frac{log\ P_{out}^{TDMA}(M, N, \eta)}{log\ (\eta)} = N. \tag{32}$$

The above analysis shows that full diversity order $N$ that is achieved by greedy opportunistic scheduling is also achievable by fixed TDMA. Considering that the two scheduling schemes are two extremes of all scheduling policies, we can infer that full diversity order can always be achieved in relay-assisted networks, as long as that the relay is selected optimally according to (7). This is a good indication that being greedy does not provide much gain in relay-assisted systems. The simple fixed TDMA scheme can achieve good system reliability and user fairness at the same time.

*B. Power gap between fixed TDMA and greedy opportunistic scheduling*

Although fixed TDMA achieves the same diversity order as greedy opportunistic scheduling, there may exist a power gap between the two scheduling schemes. Here, we quantify this power gap under symmetric channel condition, where $\Omega_{u,r} = \Omega_{r,B} = \sigma$. It will be shown in the next section that this power gap can be closed by introducing limited opportunism to TDMA.

With $\Omega_{u,r} = \Omega_{r,B} = \sigma$, the outage probability of fixed TDMA becomes

$$P_{out}^{TDMA}(N,\eta) = \prod_{r=1}^{N}\left\{1 - \exp\left\{-\frac{\tau}{\eta}\left(\frac{1}{\alpha\sigma} + \frac{1}{(1-\alpha)\sigma}\right)\right\}\right\} \\ = \left\{1 - \exp\left\{-\frac{1}{\alpha(1-\alpha)}\frac{\tau}{\eta\sigma}\right\}\right\}^{N}. \tag{33}$$

Meanwhile, the outage probability lower bound in (24) obtained by opportunistic scheduling becomes

$$\underline{P_{out}}(N,\eta) = \left\{1 - \exp\left\{-\frac{1}{1-\alpha}\frac{\tau}{\eta\sigma}\right\}\right\}^{N}. \tag{34}$$

Comparing (33) and (34), we notice that there exists a power gap of $10\log_{10}\frac{1}{\alpha}$ dB between fixed TDMA and greedy opportunistic scheduling. Recall that $\alpha \in (0,1)$ is defined in Section III.B as the portion of total power allocated to users to transmit one symbol. The power gap diminishes as we allocate more power to the users. In particular, when $\alpha = 0.5$, the power gap is 3 dB.

## V. RELAXED-TDMA SCHEDULING

In this section, we show that the power gap shown in the last paragraph can be closed by a simple relaxed TDMA scheduling scheme. That is, the simple scheme can achieve the optimal outage probability without suffering the drawbacks of greedy opportunistic scheduling. In other words, there is no longer a tradeoff between efficiency, fairness, and implementation complexity in relay assisted networks.

## A. Relaxed-TDMA Scheduling

While greedy opportunistic scheduling has full freedom to swap the transmission order of mobile users, fixed TDMA has zero. In between, we define a $k$-user relaxed-TDMA scheme, where $M$ users are divided into groups of $k$ users, i.e. $\frac{M}{k}$ groups. Then, fixed TDMA is adopted to allocate different time slots to different groups in a static manner. Within each slot, we are free to choose, among the $k$ users that are pre-assigned to the slot, the one with the highest instantaneous channel gain to transmit. By doing so, we only have small-scale opportunistic scheduling within each group. Note that greedy opportunistic scheduling and fixed TDMA are the special cases with $k = M$ and $k = 1$, respectively.

The following theorem proves that the outage probability lower bound in (24) achieved by greedy opportunistic scheduling can also be achieved by introducing only very little opportunism to fixed TDMA, i.e. increasing $k$ from 1 to 2.

**Theorem 2**: A two-user relaxed-TDMA scheme, i.e. $k = 2$, achieves the optimal outage probability in (24) at the high SNR region.

*Proof:* For a group of two users, the outage probability at the high SNR region can be obtained by letting $M = 2$ in (27). The equation shows that this outage probability only depends on the R-B channels regardless of which users are in the group. Hence, it is the same for all groups, and consequently is equal to the system outage probability. That is, the outage probability of the two-user relaxed TDMA scheme, denoted by $P_{out}^{2-user}$, is given by

$$P_{out}^{2-user} = \prod_{r=1}^{N} \left\{ \frac{\tau}{(1-\alpha)\Omega_{r,B}} \cdot \frac{1}{\eta} \right\}. \tag{35}$$

Meanwhile, at the high SNR region, $\underline{P_{out}}$ in (24) becomes

$$\begin{aligned}\underline{P_{out}}(N,\eta) &= \prod_{r=1}^{N} \left\{ 1 - \exp\left\{ -\frac{\tau}{(1-\alpha)\eta\Omega_{r,B}} \right\} \right\} \\ &\approx \prod_{r=1}^{N} \left\{ \frac{\tau}{(1-\alpha)\Omega_{r,B}} \cdot \frac{1}{\eta} \right\},\end{aligned} \tag{36}$$

which is exactly the same as $P_{out}^{2-user}$ in (1). This completes the proof. ∎

Theorem 2 proves that the two-user relaxed-TDMA scheme suffers no performance loss compared with greedy opportunistic scheduling at high SNR region. At the same time, it largely decreases computational complexity and signaling overhead, as the system only needs to estimate the U-R channels of two users and select the "better" user between the two in each time slot. We also note that the proof in Theorem 2 does not depend on a specific way of grouping users. Indeed, the theorem holds regardless of how we group the users. To implement the relaxed-TDMA scheme, a fully distributed protocol is presented in the Appendix.

## B. Relaxed-TDMA enhances fairness

Intuitively, the two-user relaxed-TDMA scheme also yields better fairness among the users compared with greedy opportunistic scheduling. This is illustrated in this subsection by the variance of channel access delay and the Jain's fairness index.

*1) Variance of channel access delay:* The delay a user experiences between two consecutive transmissions, referred to as channel access delay, is a direct reflection of quality of service received by the users. While average channel access delay is a good indicator of throughput performance, the variance of channel access delay reflects the dispersion among transmission opportunities perceived by different users.

For simplicity, suppose that all users are homogeneous, i.e., $\Omega_{u,r}$ are the same for all $u$. By symmetry, each user transmits with probability $\frac{1}{k}$ in its designated time slot. Let $T$ be the the channel access delay and $\delta$ be the length of a relay cycle. Then, the average channel access delay

$$E[T] = \sum_{i=1}^{\infty} \left(i \cdot \frac{M}{k} \delta\right) \cdot p(1-p)^{i-1} = M\delta, \tag{37}$$

and its second moment is

$$E[T^2] = \sum_{i=1}^{\infty} \left(i \cdot \frac{M}{k} \delta\right)^2 \cdot p(1-p)^{i-1} = \left(2 - \frac{1}{k}\right) M^2 \delta^2. \tag{38}$$

Hence, the delay variance is

$$Var[T] = E[T^2] - E[T]^2 = \left(1 - \frac{1}{k}\right) M^2 \delta^2, \quad 1 \leq k \leq M. \tag{39}$$

From (37) and (39), we can see that while the average channel access delay remains constant, the variance increases with the increase of group size $k$. For fixed TDMA where $k = 1$, delay variance is zero. On the other hand, the largest delay variance occurs when greedy opportunistic scheduling is adopted, i.e., $k = M$. This implies that although homogeneous users equally share the wireless resource in the long run, some of them may temporarily be in severe starvation, which leads to poor short term fairness. In contrast, with $k = 2$, two-user relaxed-TDMA largely reduces the delay variance and enhances the short term fairness among users compared with greedy opportunistic scheduling.

*2) Jain's Fairness Index:* Now, let us remove the homogeneity assumption and consider a general case where $\Omega_{u,r}$'s can be different for different $u$. Here, we use Jain's fairness index (*FI*) [25] to quantify the fairness of airtime allocation as a result of user scheduling. We show that the fairness improves as the group size $k$ decreases.

Suppose that $M$ mobile users contend for a total airtime of length $L$. Let $x_u(L)$ denote the portion of airtime received by the $u^{th}$ user during the time period $L$. The $FI$ of an airtime allocation vector $\boldsymbol{x}(L)$ is defined as

$$FI(\boldsymbol{x}(L)) = \frac{(\sum_{u=1}^{M} x_u(L))^2}{M \sum_{u=1}^{M} x_u^2(L)}. \tag{40}$$

For example, when $\boldsymbol{x}(L) = [1/3, 1/3, 1/3, 0]$, $FI(\boldsymbol{x}(L)) = 0.75$.

Jain's $FI$ is continuous and bounded between $[0,1]$. The higher the index, the fairer the airtime allocation. If mobile users equally share the airtime, i.e., $x_i(L)$'s are equal, then the fairness index is 1. On the contrary, the fairness index value tends to be low if the airtime allocation is in favor of few users.

Here, we examine the upper and lower bounds of the Jain's $FI$ of the relaxed-TDMA scheme. In particular, the upper bound occurs when the users in each group equally share the airtime, leading to $FI = 1$ regardless of the group size. Meanwhile, $FI$ reaches the lower bound when there exists a dominant user in each group that takes up all the airtime allocated to the group. The resource allocation vector becomes $\boldsymbol{x}(L) = [\underbrace{k/M, \ldots, k/M}_{M/k\ s}, \underbrace{0, \ldots, 0}_{M-M/k\ s}]$. This extremely unfair airtime allocation yields the lowest $FI$ given by

$$\underline{FI}(k) = \frac{1}{M \cdot \frac{M}{k} \cdot \left(\frac{k}{M}\right)^2} = \frac{1}{k}, \tag{41}$$

where $1 \leq k \leq M$.

We can see that the $FI$ lower bound is a function of the group size k. It overlaps with the upper bound when $k = 1$, which corresponds to fixed TDMA. Meanwhile, the minimum $FI$ lower bound occurs when $k = M$, which corresponds to greedy opportunistic scheduling. In general, the lower bounds improves as we decrease $k$. In particular, by letting $k = 2$, the two-user relaxed-TDMA guarantees that $FI$ is no less than $\frac{1}{2}$. The increase in the lower bound is a good indicator that relaxed-TDMA that limits the scale of opportunism helps to enhance user fairness, as it reduces the potential disparity of airtime allocation among users especially when $M$ is large.

## VI. SIMULATION RESULTS

In this section, we verify the analysis in the previous sections through numerical simulations. In our simulations, $N_0 = 1$ and decoding threshold $\tau = 3$. The length of a relay cycle is $2ms$ and Doppler spread is $15 Hz$. Unless otherwise specified, power is equally allocated between users and the relay set, i.e. $\alpha = 0.5$.

*A. Outage probability*

We consider a system with 8 users and 5 relays in Fig.1. The mean channel gains of R-B links are $\Omega_{r,B} = [1.2, 0.6, 0.5, 1.3, 0.7]$ and the mean channel gains of U-R links are listed in Table I. Outage probability is plotted against $P_0/N_0$ for both fixed TDMA scheduling (i.e., $k = 1$) and greedy opportunistic scheduling (i.e., $k = 8$). Both analytical and simulation results are presented. The curves that represent analytical results are calculated according to (23). It can be seen that the analytical and simulation results are on top of each other. Besides, we observe that fixed TDMA and greedy opportunistic scheduling have the same diversity gain, which validates our claim that fixed TDMA achieves optimal diversity gain in Section IV.A.

Assuming that the channels are symmetric, i.e. $\Omega_{u,r} = \Omega_{r,B} = 1$, we show in Fig. 2 that the power gap between fixed TDMA and greedy opportunistic scheduling under different power allocation parameter $\alpha$. Same as we observe in Fig. 1, fixed TDMA achieves the same diversity order as greedy opportunistic scheduling. When $\alpha = 0.5$, there is a 3 dB power gap. Moreover, the power gap decreases from 3 dB to 1 dB as we allocate more power to the users ($\alpha$ changes from 0.5 to 0.8). This verifies our analysis in Section IV.B that the power gap is $10 \log \left(\frac{1}{\alpha}\right)$ dB.

Fig. 3 depicts the system optimal outage probability as a function of $M$ and $N$. The mean channel gains are uniformly distributed in (0.5,1.5) for both U-R and R-B links. We notice that when the number of relays remains unchanged, the diversity orders remains constant, despite that the outage probability decreases as $M$ increases. On the other hand, the diversity order is increased when the number of relays increases from 5 to 8. This validates our analysis that the diversity order depends only on the number of relays.

In Fig. 4, we examine the system outage probability of $k$-user relaxed-TDMA as a function of $k$. There are 6 relays and 8 mobile users in the system, where the R-B mean channel gains are within (1.5, 2). Besides, 4 out of 8 users are located closer to the relays. Considered as "fortunate" users, their U-R mean channel gains are within (1.5,2). The other 4 "unfortunate" users are located at the edge of the cell. Their U-R mean channel gains are within (0.5, 1). We set the group size to be $k = \{1,2,4,8\}$, respectively, and the grouping pattern is random. The curves in the figure are the average performance of 100 independent grouping patterns. The figure shows that the two-user relaxed-TDMA scheme significantly decreases the system average outage probability compared with fixed TDMA scheduling. In fact, it overlaps with the outage probability lower bound at high SNR region, which verifies our analysis in Theorem 2. However, the improvement becomes marginal if we further increase the group size.

*B. Fairness improvement of relaxed-TDMA scheduling*

Fig. 5 shows the improvement of short term fairness by $k$-user relaxed-TDMA when users are homogeneous. Here Jain's fairness index is plotted against normalized Doppler frequency. The normalized Doppler frequency is Doppler frequency ($15Hz$) normalized with respect to the symbol rate ($500Hz$). In our case, one unit in X-axis represents the length of $500/15 \approx 33$ relay cycles. By rule of thumb, the fairness performance within a sliding window that is less than $10$ units can be regarded as the measurement of short term fairness, otherwise it is long term fairness. Due to homogeneity of users, each user has an equal chance to transmit in its designated time slot. Thus, $FI = 1$ in the long run regardless of the group size. Nonetheless, the figure shows that short term fairness is enhanced by having a smaller $k$ in the $k$-user relaxed-TDMA scheme ($k = 2$) compared with greedy opportunistic scheduling ($k = 8$). This verifies our access delay analysis that relaxed-TDMA enhances short term fairness in Section V.B.

When users are no longer homogeneous, Fig. 6 compares the fairness performance of greedy opportunistic scheduling and two-user relaxed-TDMA, where the grouping is random. The system model is the same as that in Fig. 4 and SNR is fixed at $15dB$. Compared with opportunistic scheduling, two-user relaxed-TDMA enhances long term fairness from about $0.6$ to $0.8$. The improvement is especially significant when the sliding window is short.

Intuitively, airtime allocation between the two users in a group is closely related to their U-R channel statistics. Fig. 7 examines the impact of grouping pattern to fairness. Three kinds of grouping patterns are compared. The "good" grouping pattern puts users with similar channel gains in a group.. The "bad" grouping pattern, however, groups users with very distinct channel gains together. The third grouping method is random grouping. We can see that grouping users who are statistically similar obtains much better fairness. Its Jain's fairness index is about 20% higher than that of random grouping.

## VII. CONCLUSIONS

In this paper, we found that the conventional tradeoff between transmission reliability and user fairness is not necessary in relay-assisted systems. With optimal relay selection, fixed TDMA achieves the optimal diversity gain that is typically only obtainable by greedy opportunistic scheduling in conventional wireless systems. In addition, by introducing limited opportunism into fixed TDMA, a simple two-user relaxed-TDMA scheduling scheme asymptotically achieves the optimal outage probability in high SNR region. Compared with greedy opportunistic scheduling, the two-user relaxed-TDMA scheme not only achieves the same transmission reliability, but also significantly improves user-level fairness and decreases computational complexity. As such, we can safely enjoy the advantages of both greedy opportunistic scheduling and fixed TDMA at the same time. For practical implementations, we have proposed a simple and fully distributed algorithm for relaxed-TDMA scheduling.

# VIII. APPENDIX

## DISTRIBUTED RELAXED-TDMA SCHEDULING PROTOCOL

Without loss of generality, consider a particular time slot that is assigned to a group of $k$ users, say user $1$ to user $k$. The distributed relaxed-TDMA scheduling protocol operates in the following steps, which is also illustrated in Fig. 8.

1). The $k$ users and the BS send pilot signals so that each relay can estimate its local channel information based on the pilots received. The $r^{th}$ relay, for instance, has knowledge about $\gamma_{u,r}$ for $u = 1,2,..k$ and $\gamma_{r,B}$. Then, the $r^{th}$ relay selects the user $u_r^*$ according to

$$u_r^* = \arg \max_{u=1,2,..,k} \min \{P_{user}\gamma_{u,r}, P_{relay}\gamma_{r,B}\}. \tag{42}$$

2). Relay $r$ attempts to access the wireless medium by setting a backoff timer according to a decreasing function of parameter $Y_r$, given by

$$Y_r = \min \{P_{user}\gamma_{u_r^*,r}, P_{relay}\gamma_{r,B}\}. \tag{43}$$

3). Then, the backoff timer counts down in a fixed time interval until it reaches zeros. Therefore, the relay $r^*$ with the largest $Y_{r^*}$ will expire its backoff timer first.

4). When its backoff timer counts down to zero, relay $r^*$ sends out a RTS packet, which contains the ID of its selected user $u_{r^*}^*$. Upon hearing this RTS packet, other relays will clear their timers and keep silent until the next time slot. Onthe other hand, the $k$ mobile users will compare their IDs to the received RTS packet. The selected user will identify itself and transmit its data packet, while the other $k-1$ users will keep silent.

5). Upon this point, the best user-relay pair has been selected exactly according to (13) in a distributed fashion. In the remainder of the time slot, the only active relay $r^*$ will receive a message from $u_{r^*}^*$ and forward it to the destination with full power $P_{relay}$.

6). In the next time slot, the scheduling process is repeated for the next group of $k$ users.

In the proposed protocol, the network infrastructure bear all the responsibility of user-relay selection and the mobile users are unaware of the relay selection process at all. Therefore, we can implement the algorithm using a very simple processor at the mobile users. Moreover, the proposed algorithm is bandwidth efficient. Except for a narrowband RTS packet, no other inter-node feedback signaling overhead is generated. In addition, the algorithm is fully distributed. All that a relay needs to do is to pick one "best" user from its own perspective and set a backoff timer accordingly.

TABLE I

MEAN CHANNEL GAINS OF THE U-R LINKS IN FIG.1

| $\Omega_{u,r}$ | $r=1$ | 2 | 3 | 4 | 5 |
|---|---|---|---|---|---|
| $u=1$ | 0.2 | 0.8 | 1.3 | 1.0 | 0.5 |
| 2 | 0.8 | 1.4 | 1.2 | 1.1 | 1.0 |
| 3 | 0.8 | 0.6 | 1.4 | 0.2 | 0.1 |
| 4 | 1.3 | 1.1 | 0.7 | 0.5 | 0.3 |
| 5 | 0.3 | 0.5 | 0.7 | 1.2 | 1.4 |
| 6 | 0.5 | 0.6 | 0.9 | 1.0 | 1.1 |
| 7 | 0.8 | 0.7 | 0.6 | 0.9 | 0.4 |
| 8 | 1.3 | 1.0 | 0.7 | 0.6 | 0.4 |

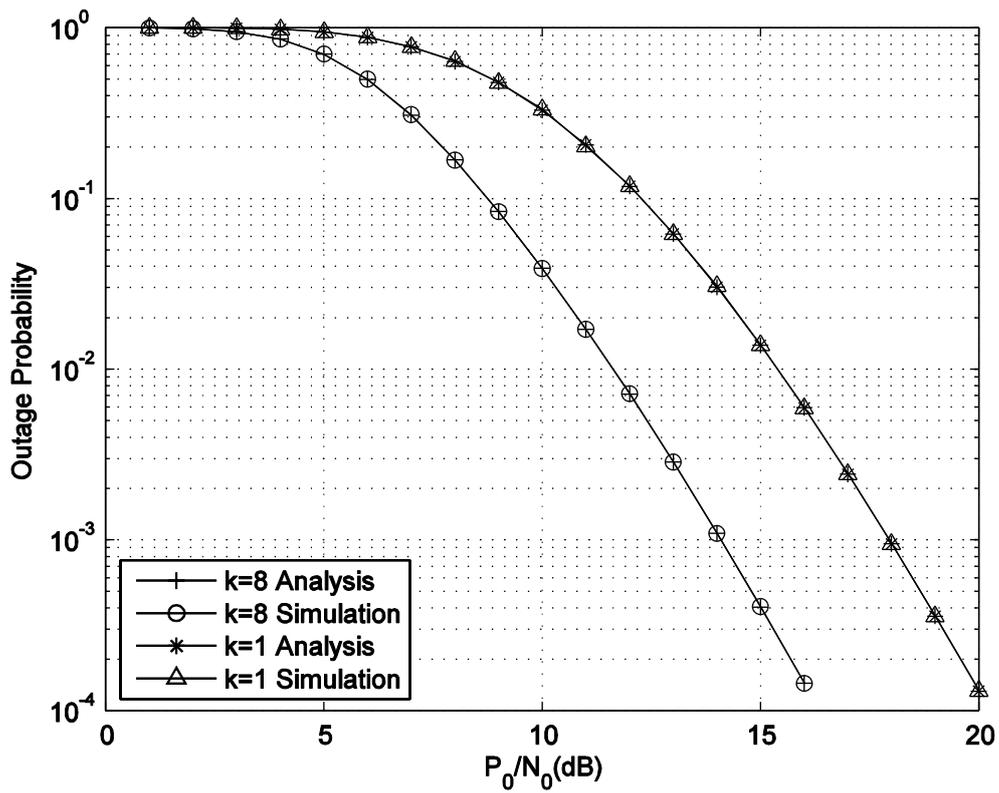

Fig.1. Analytical and simulation results of outage probability

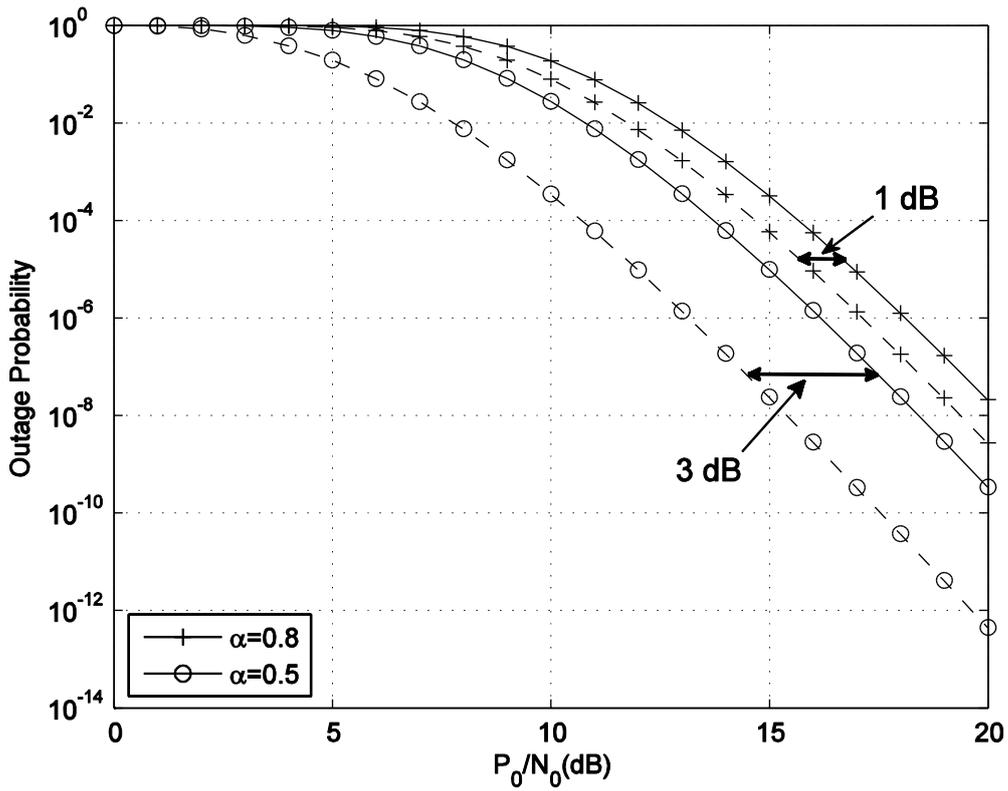

Fig.2. Power gap between fixed TDMA and greedy opportunistic scheduling

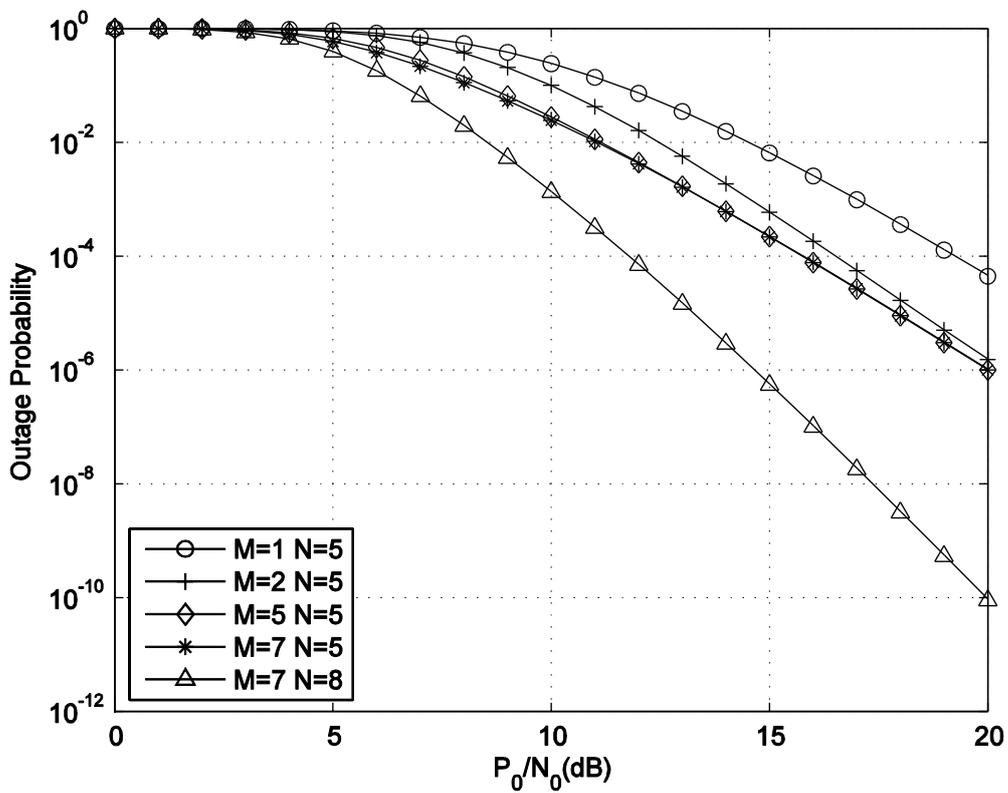

Fig.3. Outage probability of greedy opportunistic scheduling as a function of $M$ and $N$

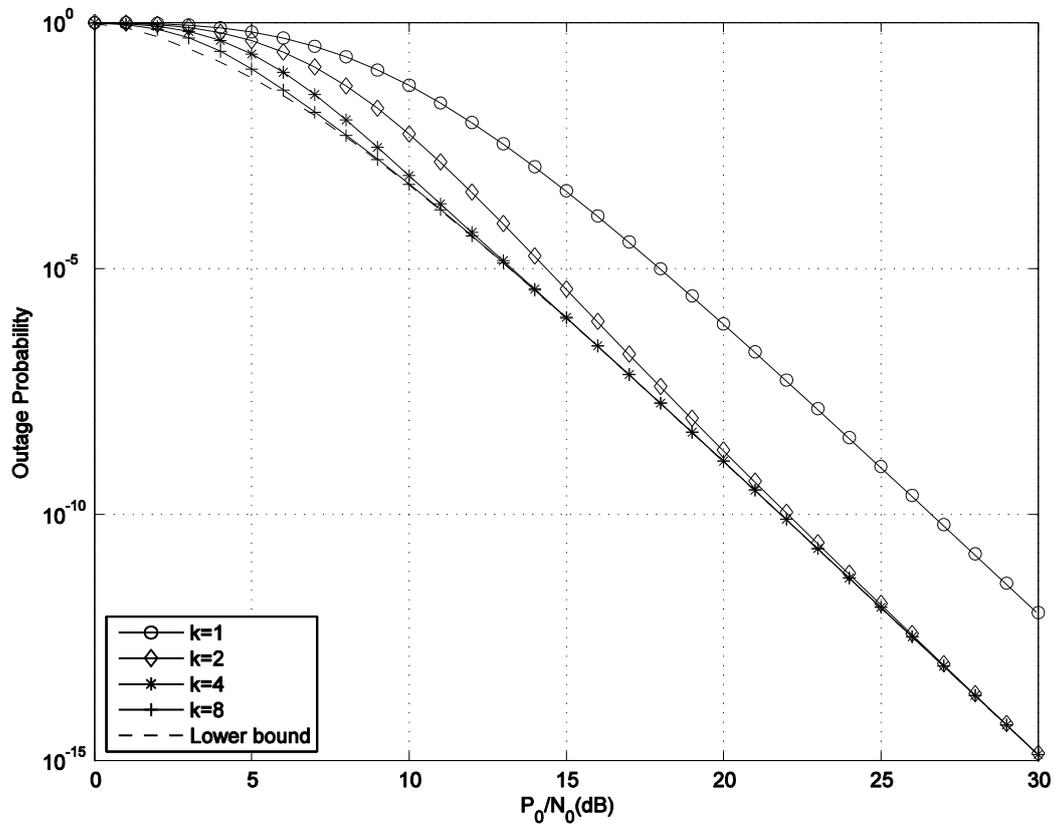

Fig.4. Outage probability of relaxed-TDMA as a function of group size $k$ ($M = 8$)

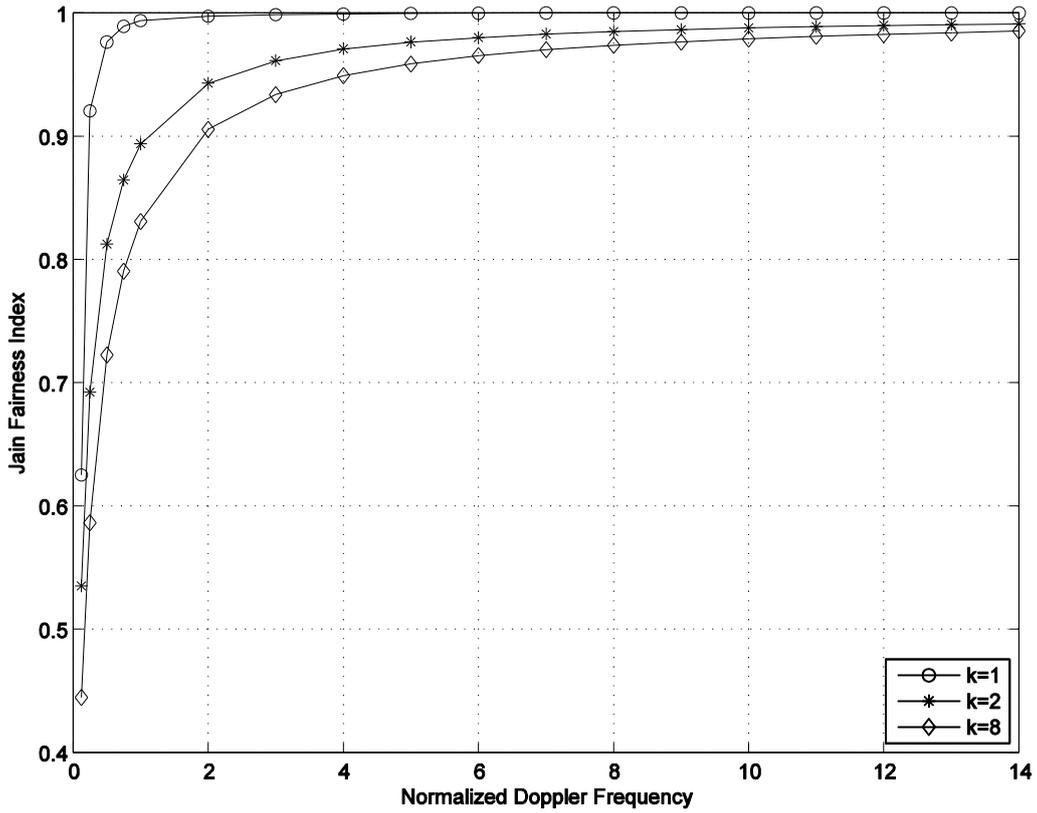

Fig.5. relaxed-TDMA improves short term fairness when users are homogeneous

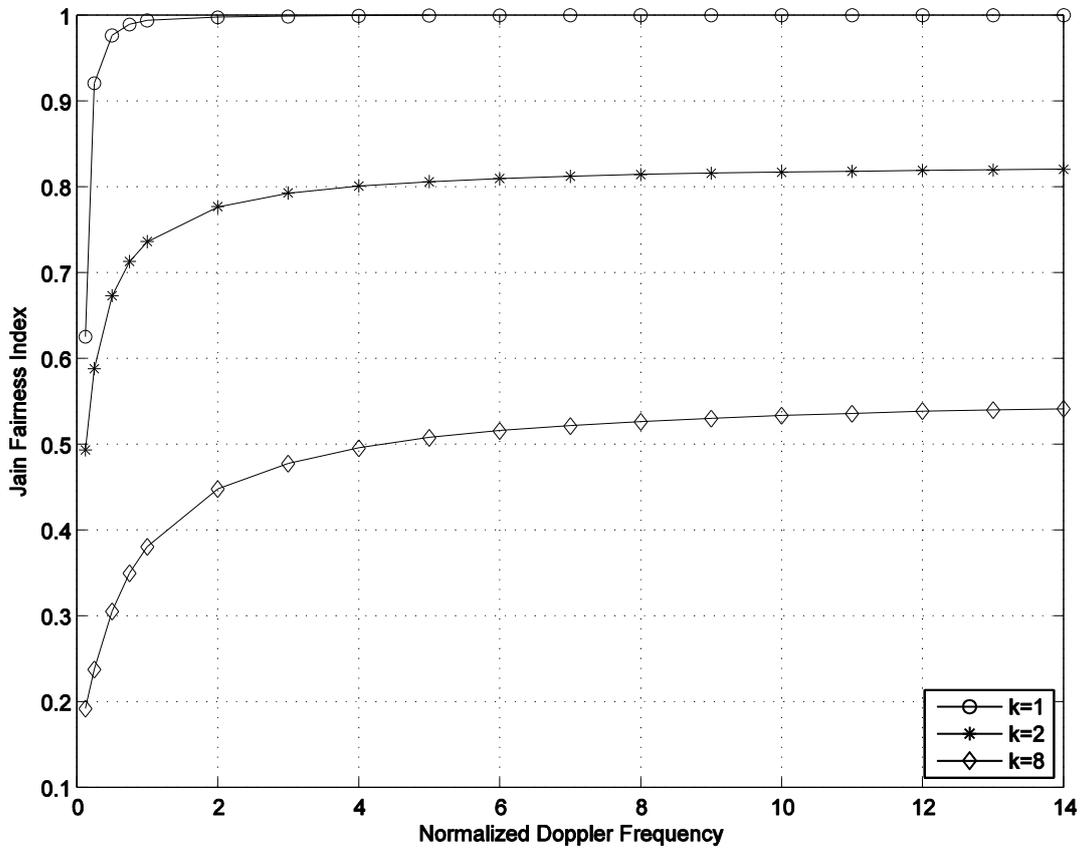

Fig.6. Fairness improvement of relaxed-TDMA under general U-R channels

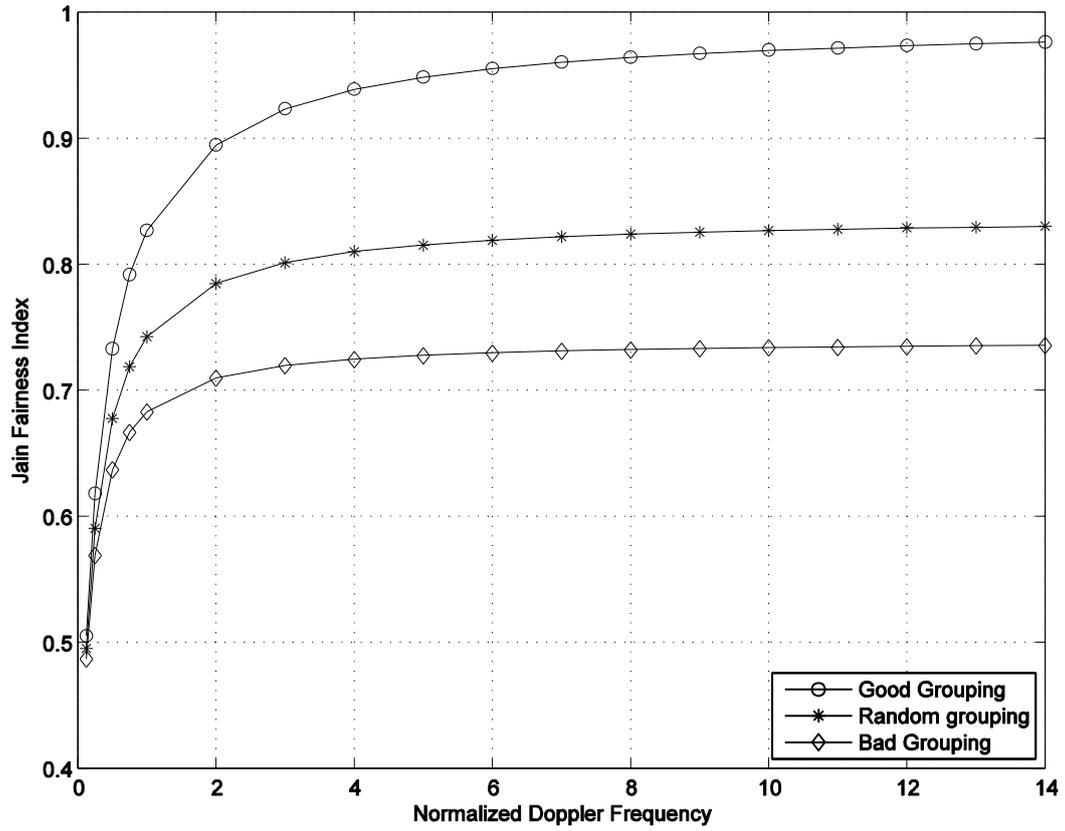

Fig.7. Impact of grouping pattern of relaxed-TDMA to fairness

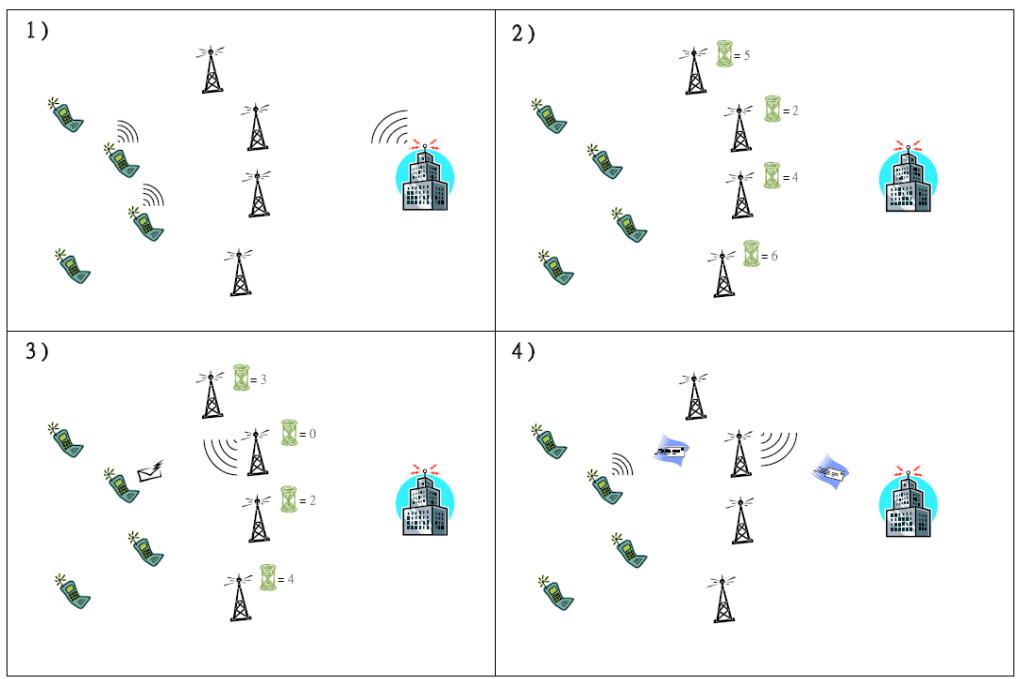

Fig.8. Distributed implementation of relaxed-TDMA scheduling